\documentclass[runningheads]{llncs}
\usepackage{hyperref}
\usepackage[T1]{fontenc}
\usepackage{graphicx}
\usepackage{mathtools}
\usepackage{amssymb}
\usepackage{todonotes}
\usepackage{multirow}
\usepackage{enumitem}
\usepackage{booktabs}
\usepackage{svg}
\usepackage{subcaption}

\newcommand{\schemename}{\textsc{ReDash}}
\newcommand{\dash}{\textsc{Dash}}

\begin{document}
\title{ReDASH: Fast and efficient Scaling in Arithmetic Garbled Circuits for Secure Outsourced Inference}

\titlerunning{ReDASH: Fast and efficient Scaling in Arithmetic Garbled Circuits}
\author{Felix Maurer\orcidID{0000-0003-4964-2103} \and
Jonas Sander\orcidID{0009-0007-9402-823X} \and
Thomas Eisenbarth\orcidID{0000-0003-1116-6973}}
\authorrunning{F. Maurer et al.}
\institute{University of Luebeck, Luebeck, Germany\\
\email{\{felix.maurer,j.sander,thomas.eisenbarth\}@uni-luebeck.de}}
\maketitle              %

\begin{center}
\small\itshape
This is an extended version of a paper accepted at SIMLA’25.
\end{center}

\begin{abstract}
\schemename{} extends \dash's arithmetic garbled circuits to provide a more flexible and efficient framework for secure outsourced inference. By introducing a novel garbled scaling gadget based on a generalized base extension for the residue number system, \schemename{} removes \dash's limitation of scaling exclusively by powers of two. This enables arbitrary scaling factors drawn from the residue number system's modular base, allowing for tailored quantization schemes and more efficient model evaluation.

Through the new $\operatorname{ScaleQuant}^+$ quantization mechanism, \schemename{} supports optimized modular bases that can significantly reduce the overhead of arithmetic operations during convolutional neural network inference. \schemename{} achieves up to a 33-fold speedup in overall inference time compared to \dash. Despite these enhancements, \schemename{} preserves the robust security guarantees of arithmetic garbling. By delivering both performance gains and quantization flexibility, \schemename{} expands the practicality of garbled convolutional neural network inference.

\end{abstract}

\section{Introduction}

As the adoption of privacy-preserving machine learning continues to grow, efficient solutions for \textit{secure outsourced inference} (SOI) remain crucial. From personalized healthcare to confidential financial analytics, organizations increasingly rely on outsourcing resource-intensive neural network computations to powerful but potentially untrusted cloud providers. Yao's \textit{garbled circuits} (GCs) \cite{DBLP:conf/focs/Yao86} ensure privacy in principle, yet their bitwise approach to arithmetic is often disadvantageous for neural network operations. In contrast, \textit{arithmetic garbled circuits} (AGCs) natively handle arithmetic computations over finite rings, making them far more efficient for deep learning tasks.

Recently, Sander et al. \cite{DBLP:journals/tches/SanderBBE25} presented the framework \dash{} which purely leverages AGCs to protect convolutional neural network inference. \dash{} provides new optimized scaling operations in the \textit{residue number system} (RNS) and novel memory layouts called LabelTensors for efficient parallelization, leading to substantial performance improvements. However, their approach is constrained by scaling limited to power-of-two factors, restricting flexibility and limiting practical performance.

We present \schemename, an extension to \dash{}, which removes its power-of-two restriction by adopting a generalized \textit{base extension} (BE) algorithm for scaling. \schemename{} allows using arbitrary scaling factors from the RNS base, enabling $\operatorname{ScaleQuant}^+$, a more efficient and flexible quantization method. The result is a significant speedup in secure inference, as \schemename{} eliminates the need to chain multiple scaling gadgets. In our evaluation, we demonstrate the large performance gain achievable through \schemename.

\section{Preliminaries}

\subsection{Secure Outsourced Inference}
Compared to \dash{}, we focus on a scenario without a TEE and only two parties (the classical GC scenario). 
We consider a compute provider who whishes to provide their computational resources to a customer while preserving the customer's data privacy. 
In \schemename{} the compute provider takes the role of the evaluator and the customer takes the role of the garbler. 
The garbler garbles the model to be outsourced during an input-independent offline phase and sends it to the evaluator. 
When the customer wants to leverage the outsourced model to perform inference, he uses an Oblivious Transfer to send the garbled inputs to the evaluator. 
Finally, the evaluator performs the secure inference and sends the garbled results back to the customer.

\subsection{Number Systems}
\noindent\textbf{Residue number systems} (RNS) represent integers using a set of pairwise co-prime moduli \(p_1, p_2, \dots, p_k\). 
A number \(x\) is represented as its residues modulo a base of moduli: $x \mapsto ([x]_1,...,[x]_k)$ where $[x]_i := x \bmod p_i$. 
An RNS with base \(p_1, p_2, \dots, p_k\) has cardinality $P_k := \prod_{i=1}^k p_i$. If $P_k$ is the product of the first $k$ prime numbers, it is referred to as the $k$-th \textit{composite primal modulus} (CPM).
The use of RNS enables efficient parallel arithmetic and simplifies the implementation of arithmetic operations in AGCs.

\noindent\textbf{Mixed radix systems} (MRS) represent an integer $x$ using position-dependent radices $r_i$: $x = d_1 + d_2 r_1 + d_3 r_1 r_2 + \dots + d_k \prod_{i=1}^{k-1} r_i$,
where each digit \(d_i\) is non-negative and smaller than it's corresponding radix $r_i$.
If for a given RNS with base $(p_1,...,p_k)$ it holds that $\forall i: r_i = p_i$ the MRS is called the \textit{associated} MRS of that RNS. 
An associated MRS has the same cardinality as the corresponding RNS.
    
\subsection{Arithmetic Garbled Circuits}
AGCs generalize Yao’s original GCs \cite{DBLP:conf/focs/Yao86} to arithmetic computations over finite rings \( \mathbb{Z}_{p} \).
Introduced by Ball et al. \cite{DBLP:conf/ccs/BallMR16}, arithmetic garbling enables efficient, secure computation of arithmetic operations. 
In AGCs, labels consist of components from $\mathbb{Z}_p$ and have corresponding semantics. 
For an input value \(a \in \mathbb{Z}_p\), the associated wire label is computed as: $l_a = l_0 + a \cdot R$, where \(l_0\) is a random base label and \(R\) is a random offset label used consistently across the circuit.

Arithmetic GCs support the following efficient operations. 
\textit{Addition gates} are ciphertext-free, where output labels are computed directly from input labels $l^{a+b} = l^a + l^b \bmod p$. 
\textit{Multiplication by a public constant} \( c \) is also ciphertext-free: $l^{ac} = c \cdot l^a \bmod p$. 
\textit{Projection gates} for unary functions \(\phi: \mathbb{Z}_p \rightarrow \mathbb{Z}_q\) require $p$ (or $p-1$ with the row-reduction optimization \cite{DBLP:journals/iacr/BallCMRS19}) ciphertexts to securely evaluate arbitrary unary functions. 
Subsequently, Ball et al. \cite{DBLP:journals/iacr/BallCMRS19} introduced an efficient mixed-modulus Half Gate that enables the multiplication of two private (non-RNS) values and an approximated sign gadget over private values in RNS representation. 
Utilizing these gadgets, ANN layers such as ReLU and MaxPooling can be efficiently constructed.

\subsection{Encoding \& Quantization}
\dash{} and \schemename{} operate on integers within a finite ring \(\mathbb{Z}_{P_k}\). 
To encode integers, positive numbers are mapped to the lower half of \(\mathbb{Z}_{P_k}\), while negative numbers are mapped to the upper half, e.g., $(0, 1, 2, -2, -1) \mapsto (0, 1, 2, 3, 4)$. 
\dash{} supports the following two quantization schemes.

\noindent\textbf{SimpleQuant} multiplies all floating-point values (weights, biases, and inputs) with a small quantization constant $\alpha$ and rounds the result to the nearest integer, where $\alpha$ balances quantization error and representable range.

\noindent\textbf{ScaleQuant} multiplies floating-point weights and inputs with a quantization constant $2^l$ and subsequently rounds to the nearest integer. 
The same procedure is applied to the bias values but using $2^{2l}$ as quantization constant. 
During inference-time, outputs of linear layers are scaled by \(2^{-\ell}\) to limit intermediate results to practical value ranges. 
The constant $l$ is chosen model dependent based on the width of the linear layers. 
$\operatorname{ScaleQuant}$ only supports quantization constants which are powers of two, as \dash's scaling gadget is limited to scaling by two.

\section{\schemename}

\subsection{Threat Model}
In this work, we adopt the classical GC scenario involving two parties: a garbler and an evaluator. 
The garbler creates the AGC and the corresponding encoding information, whereas the evaluator evaluates the garbled circuit to compute the function output. 
We assume a semi-honest security model, in which both the garbler and evaluator follow the protocol honestly but may attempt to infer additional information from the messages exchanged. 
Our scheme guarantees input privacy, output privacy, and integrity of computation under this semi-honest adversarial assumption.
\schemename{} inherits the security guarantees of \dash{} \cite{DBLP:journals/tches/SanderBBE25}, we refer to their security analysis for a detailed discussion.

\subsection{Scaling of Residue Numbers}
To support $\operatorname{ScaleQuant}$, \dash{} introduced a garbled scaling operation over the residues of RNS representations.
Scaling a number $x$ by a scaling factor $s$ means computing $y := \lfloor x/s \rfloor$.
By construction, $x = \lfloor x/s\rfloor \cdot s + (x \bmod s)$ and thus $\lfloor x/s \rfloor= (x-(x \bmod s))/s$, which is always a division with remainder zero.
We leverage that when computing $x/d \mod m$ for some divider $d$ of $x$, the equivalent operation $x d^{-1} \mod m$ can be computed on a per-residue basis as long as the inverse is well-defined for all $p_i$ of our RNS (see also \cite{szabo1967residue}).
We can thus create the division-with-remainder-zero scenario (i.e., that $s$ is a divider) by subtracting $x \mod s$ from each individual residue $x_i$ before division.
The only $s$ for which $x \mod s$ is known are the moduli of our RNS, limiting $s = p_i$ for some $i \leq k$. 
Without loss of generality, we assume that $i=k$, i.e., that $s$ is the last modulus of our RNS base.
To scale a residue $[x]_i$ down to $[y]_i = \lfloor x/s\rfloor$ for $i<k$ we compute:
\begin{equation*}
    [y]_i = ([x]_i-[x]_k) \cdot p_k^{-1}\mod{p_i}.
\end{equation*}
The equation is not well-defined for $i=k$. 
Sander et al. \cite{DBLP:journals/tches/SanderBBE25} proposed a way to determine $[y]_k$ for $s=2$ by leveraging the $\operatorname{SignGadget}$ of Ball et al. \cite{DBLP:journals/iacr/BallCMRS19} and performing a \textit{base extension} (BE) by computing $[y]_k = \operatorname{sign}([y]_2,...,[y]_{k-1}, 0)$. 
The sign-based method cannot be generalized for other choices of $s$ or for RNS bases without modulus two.
Limiting the scaling operation to $s=2$ is the main bottleneck in \dash{} when using $\operatorname{ScaleQuant}$ as it necessitates the chaining of multiple scaling layers for larger $\ell$, thereby significantly increasing the computational overhead. 
Especially in larger CNN topologies, scaling makes up a significant part of the total online runtime \cite{DBLP:journals/tches/SanderBBE25}.
We address this limitation by introducing a generalized garbled BE algorithm that enables more flexible and efficient quantization schemes trough scaling by arbitrary moduli from the RNS base.

\subsection{Generalized Garbled Base Extension}
\begin{figure}
  \centering
  \includegraphics[width=\textwidth]{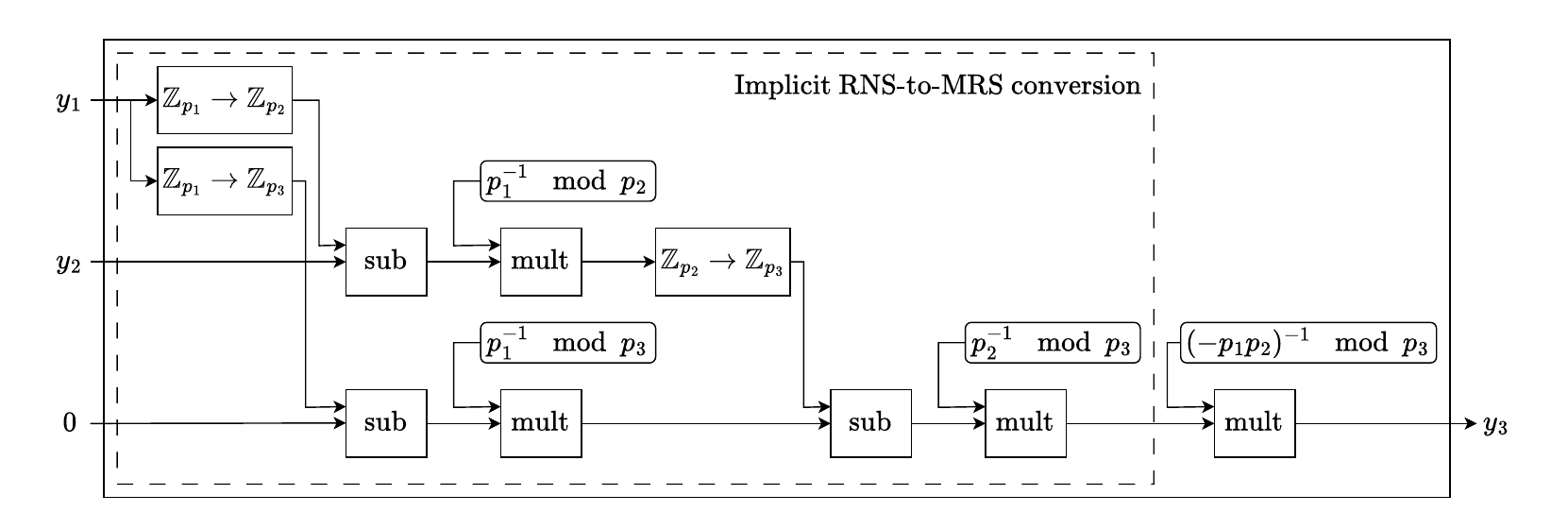}
  \caption{Circuit of the generalized BE for an RNS base of 3 moduli ($P_3$).}
  \label{fig:BEcircuit}
\end{figure}

We leverage the generalized BE detailed by Szabo and Tanaka \cite{szabo1967residue} to construct an efficient and flexible BE gadget for AGCs that allows to realize more versatile and effective quantization schemes for ANN inference in SOI.
The BE exploits the cardinality equivalence of an RNS and its associated MRS: After scaling down $x$ to $y$ by scaling factor $s=p_k$, we know from $x < P_k$ that $y < P_k/p_k = P_{k-1}$, which is the cardinality of a smaller MRS with radices $p_1,...,p_{k-1}$.
Thus, the most significant digit of $y$'s representation in its associated MRS with radices $p_1,...,p_{k}$ is not needed, hence $d_k=0$.
To determine $[y]_k$, the algorithm follows a recursive RNS-to-MRS conversion. Let $y'$ be $([y]_1,\dots, [y]_{k-1},0)$, i.e., the scaled down RNS representation before the BE. For all $j = 1,\dots,k$ we compute:
\begin{equation*}
\label{eq:rns_to_mrs}
    d_i = [z_i]_i, \text{  where }
    [z_i]_j = 
    \begin{cases}
        [y']_j,  & \text{for } i= 1, \\
        ([z_{i-1}]_j-[z_{i-1}]_{i-1})p_{i-1}^{-1} \bmod p_j,  & \text{otherwise.}
    \end{cases}
\end{equation*}
The missing residue is given by $[y]_k = -(\prod_{i=1}^{k-1} p_i)^{-1} \cdot [z_{k-1}]_k \bmod p_k$.
Figure \ref{fig:BEcircuit} depicts our generalized garbled BE for $k=3$.  
As our only goal is to determine $[y]_3$, we compute $[z_i]_j$ only for choices of $j$ that contribute to $d_3 = [z_3]_3$ and therefore contribute to computing $[y]_3$, i.e., $j=1,2$ for $i =1$ and $j=2$ for $i=2$. The garbling of this circuit is really cheap compared to previous work: All $p_i$ are stored as plaintext, thus all multiplications with $p_i^{-1}$ are free. Garbled subtractions are realized using free modular addition of labels. The garbled BE thus requires only $\sum_{i=1}^{k-1}\sum_{j=i}^{k-1}p_j$ ciphertexts per input.

\subsection{A generalized Scaling Gadget}
\begin{figure}[!t]
    \centering
    \includegraphics[width=0.4\textwidth]{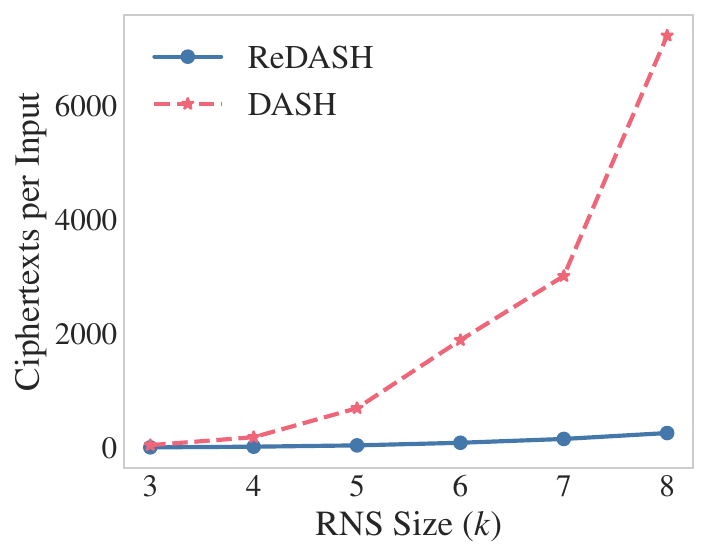}    \caption{Cost in terms of ciphertexts for garbled scaling with scaling factor $s=2$ in \schemename{}, and \dash{} with CPM RNS bases.}
    \label{fig:ciphertexts}
\end{figure}

Building upon their scaling-by-two operation, Sander et al. \cite{DBLP:journals/tches/SanderBBE25} constructed a garbled scaling gadget for \dash{}. 
The scaling gadget operates in four subsequent steps: First, a constant addition of $P_k/2$ (shift-up) ensures all values are positive. 
Then, these positive values are scaled via \dash{}'s garbled scaling-by-two operation, followed by their sign-based BE to determine $[y]_{p_1}$.
Finally, a subtraction of $P_k/4$ (shift-down) reestablishes the correct encoding of positive values to the lower half and negative values to the upper half of \(\mathbb{Z}_{P_k}\).
As part of \schemename{}, we propose a more flexible and efficient generalization of this approach, enabling scaling by arbitrary moduli of the used RNS base.
We developed a really efficient and scalable garbled implementation a BE algorithm outlined by Szabo and Tanaka. 
Where previously each $s=2$ scaling step required dedicated shift-up and shift-down operations, it suffices to shift up inputs once by $P_k/2$ before scaling and down by $P_k/2s$ after scaling. 
Inspired by the example given by Sander et al. \cite{DBLP:journals/tches/SanderBBE25}, the four steps of our generalized scaling gadget are demonstrated exemplary in \autoref{tab:scaling_example}.
Note that \schemename{}'s BE approach can, unlike \dash{}'s $\operatorname{sign}$-based solution, extend the RNS base by not just one but any number of moduli, allowing $\operatorname{ScaleQuant^+}$ to scale by a product of multiple RNS base moduli at once. \schemename{}'s scaling is (independent of $s$) cheaper in terms of ciphertexts than \dash{}'s scaling approach that leverages costly $\operatorname{sign}$-operations. Figure \ref{fig:ciphertexts} compares \dash's garbled scaling cost in terms of ciphertexts to the cost of our new garbled scaling gadget when scaling with $s=2$.

\begin{table}
    \centering
    \caption{Step-by-step outputs of \schemename{}'s generalized scaling gadget with scaling factor $s=3$ for all inputs in $\mathbb{Z}_{6}$. $x$: Input value to the scaling function. $x^\pm$: Sign information of $x$. $x^\uparrow$ and $x^\downarrow$: Output of the $\operatorname{ShiftUp}$ and $\operatorname{ShiftDown}$ operations. $b$: Scaling operation after the $\operatorname{ShiftUp}$ and before the base extension. $\varphi(x)$: Maps $x$ to its residue representation in $P_2$ base. $y$: Output of our generalized base extension algorithm.}
    \label{tab:scaling_example}
    \setlength\tabcolsep{2.2pt}
    \begin{tabular}{c|c|c|c|c|c|c|c|c|c|c|c|c|c}
    $x$ & $x^{\pm}$ & $\varphi(x)$ & $x^\uparrow$ & $x^{\pm \uparrow}$ & $\varphi(x^{\uparrow})$ & $y' = [b(\varphi(x)), 0]$ & $\varphi^{-1}(y')$ & $\varphi^{-1}(y')^\pm$ & $y$ & $\varphi(y)$ & $\varphi(y^\downarrow)$ & $y^\downarrow$ & $y^{\pm\downarrow}$ \\ \hline
    0 & 0 & [0, 0] & 3 & -3 & [1, 0] & [1, 0] & 3 & -3 & 1 & [1, 1] & [0, 0] & 0 & 0 \\
    1 & 1 & [1, 1] & 4 & -2 & [0, 1] & [1, 0] & 3 & -3 & 1 & [1, 1] & [0, 0] & 0 & 0 \\
    2 & 2 & [0, 2] & 5 & -1 & [1, 2] & [1, 0] & 3 & -3 & 1 & [1, 1] & [0, 0] & 0 & 0 \\
    3 & -3 & [1, 0] & 0 & 0 & [0, 0] & [0, 0] & 0 & 0 & 0 & [0, 0] & [1, 2] & 5 & -1 \\
    4 & -2 & [0, 1] & 1 & 1 & [1, 1] & [0, 0] & 0 & 0 & 0 & [0, 0] & [1, 2] & 5 & -1 \\
    5 & -1 & [1, 2] & 2 & 2 & [0, 2] & [0, 0] & 0 & 0 & 0 & [0, 0] & [1, 2] & 5 & -1 \\
    \end{tabular}
\end{table}

\subsection{$\operatorname{ScaleQuant^+}$}

$\operatorname{ScaleQuant}$ is limited by \dash's scaling support. 
Function-wise, $\operatorname{ScaleQuant}$ in \dash{} is restricted to quantizing with powers-of-two: $2^\ell$. 
Performance-wise, \dash{} has to chain $\ell$ scaling gadgets to perform a single down-scaling needed on the outputs of each linear layer.

We propose $\operatorname{ScaleQuant^+}$ to overcome these limitations.
$\operatorname{ScaleQuant^+}$ deploys our improved garbled scaling gadget, allowing arbitrary moduli of the RNS base as quantization constants. 
Furthermore, the $\operatorname{sign}$-gadgets used in \dash's $\operatorname{ScaleQuant}$ implementation require the used RNS base to contain modulus $p_k=2$. 
In \schemename{}, quantization via $\operatorname{ScaleQuant^+}$ allows us to omit this requirement, enabling more optimized RNS bases.
The correct quantization constant for any scheme is determined by the value range one must cover for a given inference computation. As the gap between $2^{\ell}$ and $2^{\ell+1}$ becomes increasingly large in realistic inference scenarios, the minimum required choice of $\ell$ may overshoot the scenario's value range significantly. The potential runtime performance and ciphertext overhead of suboptimal large value ranges are mitigated by leveraging $\operatorname{ScaleQuant^+}$ where RNS with more fine-grained cardinality are possible.

\subsection{Implementation Tweaks} 
\label{sec:implementation_tweaks}
We identify two optimization angles in \dash{}'s implementation of linear operations. \textit{First}, we replace the naive implementation of dense and convolutional layers on LabelTensors with BLAS-supported Eigen functions. \textit{Second}, \dash{} deploys 16-bit integers to represent garbled residues. The 16-bit value range necessitates frequent modular reduction, especially in multiplication-heavy linear layers, creating a runtime bottleneck. We leverage 32-bit data types to perform operations over residues and significantly reduce the number of expensive modular reductions. Experiments with 64-bit data types (both everywhere and in linear layers only) that necessitate even less modular reductions yielded no further performance benefit. %

\section{Evaluation}
We implemented our new scaling gadget and the $\operatorname{ScaleQuant^+}$ mechanism on top of \dash\footnote{\hyperlink{https://github.com/UzL-ITS/dash}{https://github.com/UzL-ITS/dash}}. As in the evaluation of \dash{} we focus on the time critical online phase, omitting the input-independent offline phase in our evaluation. All measurements of both \dash{} and \schemename{} were conduced on a server equipped with a single Intel Xeon Gold 5415+ CPU and a base clock of 2.90 GHz. For simplicity, we omit the online communication costs of \dash{} and \schemename{} as they are the same.

\begin{figure}
    \centering
    \includegraphics[width=\textwidth]{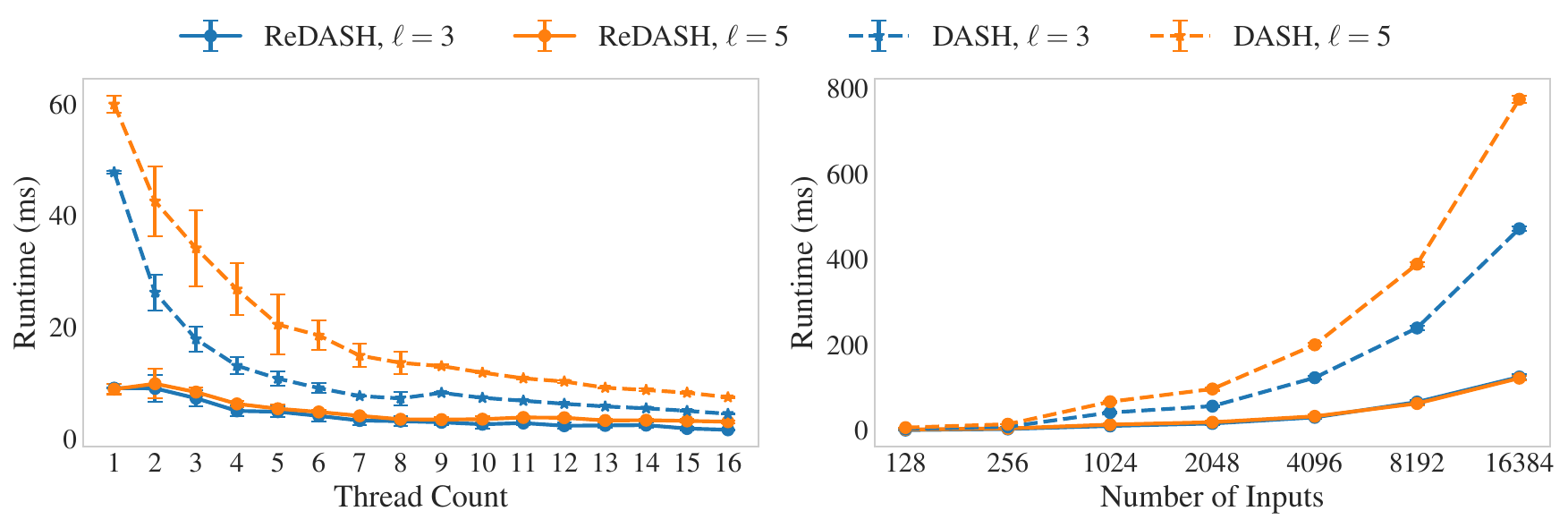}
    \caption{Online runtime comparison of the scaling gadgets in \schemename{} and \dash{}. Left shows the runtime with growing thread count and fixed input size of 128. Right shows the runtime with growing input size and a fixed number of 16 compute-threads.}
    \label{fig:microbenchmark}
\end{figure}

To evaluate the performance achieved by \schemename{}, we run an isolated micro-benchmark of our new scaling gadget. We used the CPM with the first 8 prime numbers as RNS base and replaced $p_8 = 2$ with $2^\ell$ for \schemename. \autoref{fig:microbenchmark} shows the scalability of our scaling gadget compared to \dash's scaling gadget in terms of runtime with growing numbers of compute-threads and inputs. \schemename{} beats \dash{} in both cases strictly and by a large margin.
\dash{} requires over 6 or over 11 threads to achieve the same runtime as a single-threaded \schemename{} for $\ell=3$, respective $\ell=5$. For an input of size $2^{14}$ \dash{} requires over 6x more time compared to \schemename{} to evaluate the scaling gadgets.

\begin{figure}[t]
  \centering
  \begin{subfigure}[t]{0.7\textwidth}
    \centering
    \includegraphics[width=\textwidth]{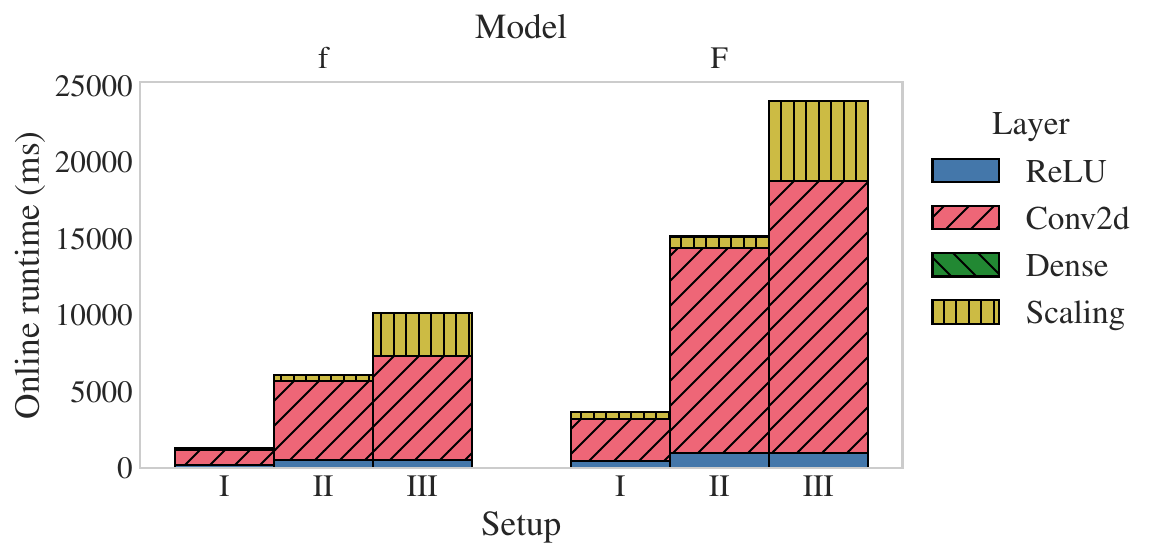}
    \caption{Comparison between 
      $\schemename{}^*$ with short RNS bases (I), \schemename{} with standard RNS bases (II) 
      and \dash{} (III).}
    \label{fig:runtime_distribution_layer_comparison}
  \end{subfigure}
  
  \vspace{1em} %
  
  \begin{subfigure}[t]{0.7\textwidth}
    \centering
    \includegraphics[width=\textwidth]{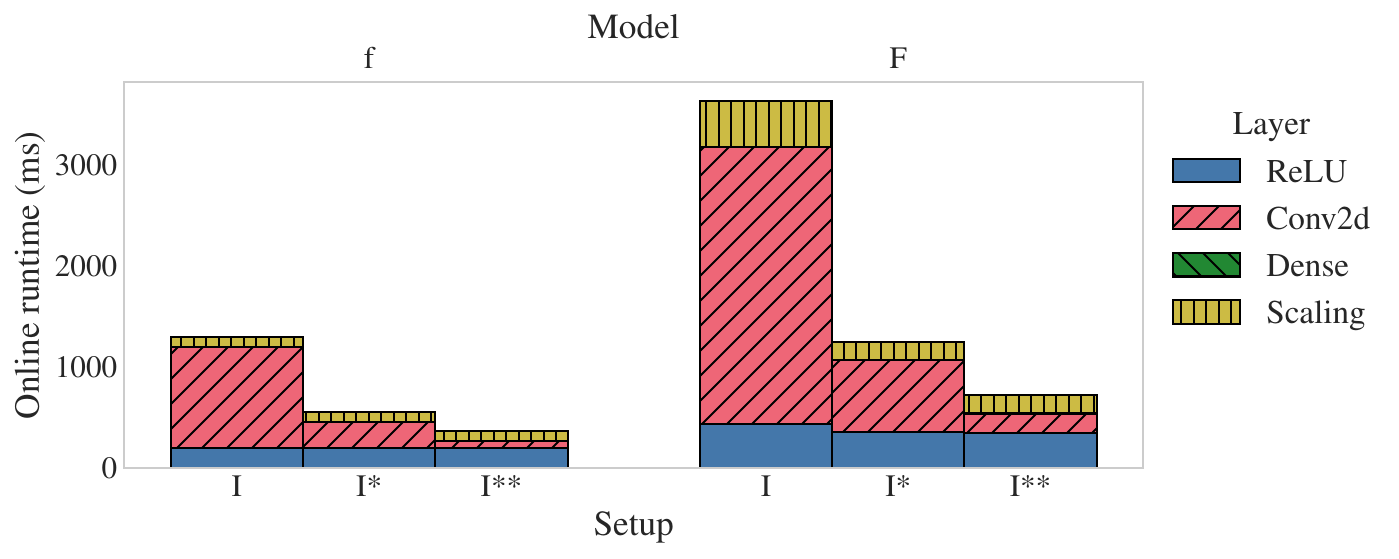}
    \caption{Comparison between 
      $\schemename{}^*$ with short RNS bases (I), Setup I with Eigen operations in linear layers (I*) 
      and Setup I with Eigen operations in linear layers and 32-bit garbled residues (I**).}
    \label{fig:runtime_distribution_layer_tweaks}
  \end{subfigure}
  
  \caption{Online inference runtimes and runtime distributions. 
    (a) Baseline comparison of \schemename{} vs.\ \dash. 
    (b) Effect of further implementation optimizations on \schemename{}. Notably, \schemename{} elimiated both the rescaling and the linear layer performance bottlenecks. }
  \label{fig:combined_runtime_distribution_layer}
\end{figure}

We compare \schemename{} to \dash{} on the two CIFAR-10 model architectures (f, F) proposed by Sander et al. \cite{DBLP:journals/tches/SanderBBE25} to be used with $\operatorname{ScaleQuant}$. Model details are listed in appendix \ref{appendix:models}. For measurements of \dash{} we used $\operatorname{ScaleQuant}$ with $\ell=5$, for \schemename{} we deployed our new quantization scheme $\operatorname{ScaleQuant^+}$ with $s=32$. We compare three different setups for $\operatorname{ScaleQuant^+}$: In the \schemename{} setup, we used the same bases as in our isolated scaling gadget benchmark. In $\schemename{}^*$, we used the following two optimized short RNS bases, which were previously not possible without $\operatorname{ScaleQuant^+}$: $(32, 167, 173)$ for model f and $(32, 97, 107)$ for model F. The final setup $\schemename{}^*_E$ extends $\schemename{}^*$ by deploying the implemenation-specific tweaks outlined in Section \ref{sec:implementation_tweaks}.
While the scaling operation of \dash{}'s $\operatorname{ScaleQuant}$ scheme covers up to 28\% of total inference runtime, our new approach reduces scaling runtime contribution by 21\% to a total of 8\% in $\schemename{}^*$ for model architecture f, as seen in Table \ref{tab:runtime_distribution}.
The total inference runtime and the relative cost of each NN operation with and without optimized bases compared to \dash{} is visualized in Figure \ref{fig:combined_runtime_distribution_layer}. 
Overall, our new quantization scheme that deploys the optimized and generalized scaling gadget is up to seven times faster in deep CNN architectures compared to \dash{}'s state-of-the-art solution.
In a second inference benchmark, we evaluate the online runtime performance impact of our improved implementation of linear operations in \schemename{}. Figure \ref{fig:combined_runtime_distribution_layer} shows a further 70-80\% speedup achieved via Eigen integration and 32-bit garbled RNS labels. When combining the new scaling features of \schemename{} with these implementation tweaks, we achieve an up to 33-fold speedup compared to \dash{}. 

\setlength{\tabcolsep}{1pt}
\begin{table}
\centering
\caption{Per-layer online runtimes (ms) for \schemename{} and \dash. 
  $\schemename^*$ means \schemename{} uses optimized short RNS bases. 
  $\schemename^*_{E}$ uses optimized short RNS bases, Eigen operations and 32-bit garbled label residues.}
\begin{tabular}{|c|cccc|cccc|}
\hline
\multirow{2}{*}{Operation}
  & \multicolumn{4}{c|}{Model f}
  & \multicolumn{4}{c|}{Model F} \\
  & \dash{} 
  & \schemename{} 
  & $\schemename^*$ 
  & $\schemename^*_{E}$
  & \dash{} 
  & \schemename{}
  & $\schemename^*$ 
  & $\schemename^*_{E}$ \\
\hline
Dense     & 0.6   & 0.5    & 0.2 &  \textbf{0.1} 
          & 0.6   & 0.6    & 0.2 &  \textbf{0.1} \\ 
Conv2d    & 6771  & 5135   & 997 &  \textbf{71} 
          & 17660 & 13393  & 2742 &  \textbf{193} \\
ReLU      & 538   & 529    & 197  &  \textbf{197} 
          & 981   & 982    & 433  &  \textbf{346} \\
Scaling   & 2841  & 437    & 99   &  \textbf{97} 
          & 5274  & 763    & 455  &  \textbf{179} \\
\hline
$\sum$    & 10150.6 & 6101.5 & 1293.2 & \textbf{365.1} 
          & 23915.6 & 15138.6 & 3630.2 & \textbf{718.1} \\
\hline
\end{tabular}
\label{tab:runtime_distribution}
\end{table}

\section{Conclusion}
We introduced \schemename, an extension of Dash that enables efficient scaling by arbitrary RNS moduli in AGCs. By replacing \dash{}’s \cite{DBLP:journals/tches/SanderBBE25} sign-based scaling with a generalized BE, \schemename{} supports more flexible quantization through $\operatorname{ScaleQuant^+}$, eliminating the restriction to power-of-two scaling.
Our evaluation shows that our solution improves runtime performance in the SOI setting by up to 33 times compared to \dash{}. These improvements make secure outsourced inference with AGCs more practical, scalable, and adaptable to real-world tasks.

\bibliographystyle{splncs04}
\bibliography{ref}

\begin{thebibliography}{1}
\providecommand{\url}[1]{\texttt{#1}}
\providecommand{\urlprefix}{URL }
\providecommand{\doi}[1]{https://doi.org/#1}

\bibitem{DBLP:journals/iacr/BallCMRS19}
Ball, M., Carmer, B., Malkin, T., Rosulek, M., Schimanski, N.: Garbled neural networks are practical. {IACR} Cryptol. ePrint Arch. p.~338 (2019), \url{https://eprint.iacr.org/2019/338}

\bibitem{DBLP:conf/ccs/BallMR16}
Ball, M., Malkin, T., Rosulek, M.: Garbling gadgets for boolean and arithmetic circuits. In: Weippl, E.R., Katzenbeisser, S., Kruegel, C., Myers, A.C., Halevi, S. (eds.) Proceedings of the 2016 {ACM} {SIGSAC} Conference on Computer and Communications Security, Vienna, Austria, October 24-28, 2016. pp. 565--577. {ACM} (2016). \doi{10.1145/2976749.2978410}, \url{https://doi.org/10.1145/2976749.2978410}

\bibitem{DBLP:journals/tches/SanderBBE25}
Sander, J., Berndt, S., Bruhns, I., Eisenbarth, T.: Dash: Accelerating distributed private convolutional neural network inference with arithmetic garbled circuits. {IACR} Trans. Cryptogr. Hardw. Embed. Syst.  \textbf{2025}(1),  420--449 (2025). \doi{10.46586/TCHES.V2025.I1.420-449}, \url{https://doi.org/10.46586/tches.v2025.i1.420-449}

\bibitem{szabo1967residue}
Szabo, N.S., Tanaka, R.I.: Residue arithmetic and its applications to computer technology. (No Title)  (1967)

\bibitem{DBLP:conf/focs/Yao86}
Yao, A.C.: How to generate and exchange secrets (extended abstract). In: 27th Annual Symposium on Foundations of Computer Science, Toronto, Canada, 27-29 October 1986. pp. 162--167. {IEEE} Computer Society (1986). \doi{10.1109/SFCS.1986.25}, \url{https://doi.org/10.1109/SFCS.1986.25}

\end{thebibliography}

\appendix
\section{Model Architectures}
\label{appendix:models}
Like Sander et al. \cite{DBLP:journals/tches/SanderBBE25} we used their models f and F for the evaluation (see \autoref{tab:model_architectures}).
\begin{table}
    \setlength{\tabcolsep}{2pt}
	\centering
	\caption{Model architectures. $R$: ReLU, $(a)$: dense layer with $a$ outputs, $(a,b,c,d)$: 2d convolution with $a$ input-channel, $b$ output-channel, filter size $c$ and a stride of $d$.}
	\label{tab:model_architectures}
 \scalebox{1}{
	\begin{tabular}{ll}
	\toprule
	Modelf: & $(3,32,3,1), R, (32,32,3,1), R, (32,32,2,2), (32,64,3,1), R, (64,64,3,1), R,$\\
      & $(64,64,2,2), (64,128,3,1), R, (128,128,3,1), R, (10)$\\
	ModelF: & $(3,64,3,1), R, (64,64,3,1), R, (64,64,2,2), (64,64,3,1), R, (64,64,3,1), R,$\\
      & $(64,64,2,2), (64,64,3,1), R, (64,64,1,1), R, (64, 16, 1, 1), R, (10)$\\
	\bottomrule  
	\end{tabular}
}
\end{table}

\end{document}